\documentclass[aps,prl,twocolumn,superscriptaddress,showpacs]{revtex4-2}
\usepackage{graphicx}
\usepackage{mathrsfs}
\usepackage{bm}
\usepackage{amsmath}
\usepackage{amsfonts}
\usepackage{multirow}
\usepackage{bigstrut}
\usepackage{rotating}
\usepackage{booktabs}
\usepackage{color}
\usepackage{braket}
\usepackage{dcolumn}
\usepackage{enumerate}
\begin{document}

\title{Engineering topologically-protected zero-dimensional interface end states in antiferromagnetic heterojunction graphene nanoflakes}
%Engineering tunability zero-dimensional topological interface states in graphene nanoribbons

\author{Cheng-Ming Miao}
\affiliation{College of Physics, Hebei Normal University, Shijiazhuang 050024, China}

\author{Yu-Hao Wan}
\affiliation{International Center for Quantum Materials, School of Physics, Peking University, Beijing 100871, China}

\author{Qing-Feng Sun}
\email[]{sunqf@pku.edu.cn}
\affiliation{International Center for Quantum Materials, School of Physics, Peking University, Beijing 100871, China}
\affiliation{Hefei National Laboratory, Hefei 230088, China}
\affiliation{CAS Center for Excellence in Topological Quantum Computation, University of Chinese Academy of Sciences, Beijing 100190, China}

\author{Ying-Tao Zhang}
\email[]{zhangyt@mail.hebtu.edu.cn}
\affiliation{College of Physics, Hebei Normal University, Shijiazhuang 050024, China}

\begin{abstract}
We investigate the energy band structure and energy levels of
a heterojunction composed of two antiferromagnetic graphene nanoflakes
with opposite in-plane antiferromagnetic orderings,
in which the modified Kane-Mele model is employed.
 Before forming an antiferromagnetic graphene heterojunction,
the energy gap of helical edge states in each isolated graphene nanoflake
are opened by the antiferromagnetic ordering
and there is no the in-gap corner state.
We find that when two opposite antiferromagnetic graphenes
are coupled to form a heterojunction nanoflake,
topologically-protected zero-dimensional in-gap states can be induced.
In addition, we demonstrate that the in-gap states locate
at the end of the interface and are robust against magnetic disorder,
Anderson disorder, and interfacial magnetic defects.
The position and number of the in-gap interface end states
in the heterojunction sample can be precise quantum controlled.
\end{abstract}
\maketitle
%========================================================================================
\section{\uppercase\expandafter{\romannumeral 1}. Introduction}
%========================================================================================
%main content
Higher-order topological insulators (HOTIs) have received great interest
and rapid development both theoretically and experimentally in the past six years \cite{Benalcazar2017, Benalcazar2017a, Song2017, Langbehn2017, Peng2017, Ezawa2018, Ezawa2018a, Schindler2018, Kunst2018, Khalaf2018, Fukui2018, Park2019, Banerjee2020, Ren2020, Liu2021a, Miao2022, Liu2022, Qian2022, Zeng2022,
addref1,Hua2023}.
An \emph{n}th-order topological insulator (TI) has protected gapless edge states of codimension $d_c=n$.
For example, a two-dimensional second-order topological system hosts zero-dimensional (0D) corner states of codimension $d_c=2$.
The conventional TIs which host codimension $d_c=1$ gapless edge states are also called first-order TIs \cite{Moore2010,Hasan2010,Qi2011,Teo2010,Essin2011,addref2,Isaev2011,Bansil2016,Chiu2016,Ren2016}.
The gapless edge states with codimension $d_c \geq 2$ have been observed in a variety of systems experimentally such as electrical circuits \cite{Imhof2018, Peterson2018, Ezawa2019a, Ezawa2019, Serra-Garcia2019, Zhang2021}, acoustic \cite{Ni2018, Xue2018, Xue2019, Xue2020, Gao2021}, photonic crystals \cite{Serra-Garcia2018, Li2019, Xie2019, Chen2019, Mittal2019, Hassan2019, Kim2020}, mechanical \cite{Wakao2020}, and solid-state materials \cite{Schindler2018a} $et~al$.
Although HOTIs in electronic systems have not yet been experimentally realized, there have been extensive theoretical explorations, including consideration of disordered \cite{Wang2020, Li2020, Yang2021}, non-Hermiticity \cite{Liu2019, Luo2019, Liu2021}, electron-electron interactions \cite{Kudo2019, Peng2020, Zhao2021}, electron-phonon interaction \cite{Lu2023}, periodic \cite{Hu2020, Huang2020, Pan2020, Ghosh2020, Nag2021, Ning2022, Ghosh2022, Ghosh2022a} or magnetic driving \cite{Ezawa2018b, Sheng2019, Ren2020, Han2022, Miao2022, Zhu2023}.

It is well known that the phase transition from first-order TIs to HOTIs needs to break the symmetry protecting first-order topological states by taking advantage of the properties of anisotropy \cite{Yan2019}. Applying an in-plane magnetic ordering to the Kane-Mele model, several works have shown that a second-order topological phase occurs hosting the topological in-gap 0D corner states \cite{Kane2005, Ren2020, Han2022, Miao2022, Zeng2022}.
However, the zigzag edges in graphene nanoribbons are one of the necessary conditions in their proposals. The reason is that the one-dimensional gapless edge states could be gapped on the zigzag edges by the in-plane ferromagnetic or antiferromagnetic ordering, but it remains the gapless edge states on the armchair edges \cite{Son2006, Tang2016, Lin2020, Miao2022, Zhu2023}. This indicates that the properties of anisotropy are necessary but insufficient conditions for realizing HOTIs.

Usually, the 0D corner states are localized at sharp corners where two boundaries form a Dirac-mass domain wall with opposite signs in a two-dimensional system \cite{Ren2020, Chen2020, Zhuang2022}. To overcome the constraints of boundary conditions, Yang $et~al$. have proposed domain wall-induced topological corner states in the coexistence of sublattice symmetry breaking and lattice deformation in an artificial graphene system \cite{Yang2020}. Zhu $et~al$. have predicted the emergence of topological corner states at a sensitive sublattice dependence domain wall \cite{Zhu2022, Zhu2023}. However, these proposals are still heavily dependent on the choice of boundaries.

In this work, we consider a heterojunction composed of two modified
Kane-Mele graphene nanoflakes with opposite in-plane antiferromagnetic
orderings, as shown in Fig.~\ref{fig1},
where the zigzag (armchair) edge is along the $y$ ($x$)-axis.
The magnetic ordering directions of the sublattice sites $A$ and $B$
in the blue (green) region are respectively along the $+x$($-x$)-axis
and $-x$($+x$)-axis,
and two graphene nanoflakes with opposite antiferromagnetic orderings
are coupled to form a heterojunction.
For each isolated antiferromagnetic graphene nanoflake,
the helical edge states open the gap by
the antiferromagnetic ordering and there is no in-gap corner state.
We find that when they are coupled to form a heterojunction nanoflake,
topologically-protected 0D in-gap states can be induced
by the formation of the heterojunction and they locate at the
end of the interface.
These 0D in-gap interface end states are very
robust against magnetic disorder, Anderson disorder,
and interfacial magnetic defects.
The origin of the 0D in-gap states is explained by using
the effective mass terms of the helical edge states.
In addition, we also study other heterojunction configurations
and irregular coupling boundaries.
The results show that multiple topologically-protected 0D in-gap
states emerge at the sample interface
and are tunable independent of boundary conditions.

The rest of the paper is organized as follows.
In Sec. II, we derive the tight-binding Hamiltonian for
the heterojunction composed of two graphene
nanoflakes with opposite in-plane antiferromagnetic orderings.
In Sec. III A, we show the band structure and energy levels of the
heterojunction and demonstrate the emergence of
the topologically-protected 0D in-gap interface end states.
Then in Sec. III B-D, we verify the robustness, show the origin,
and discuss the tunability of the 0D in-gap interface end states,
respectively.
Finally, a summary is presented in Sec. IV.

%========================================================================================
\section{\uppercase\expandafter{\romannumeral 2}. model and hamiltonian}
%========================================================================================
\begin{figure}
	\centering
	\includegraphics[width=8.6cm,angle=0]{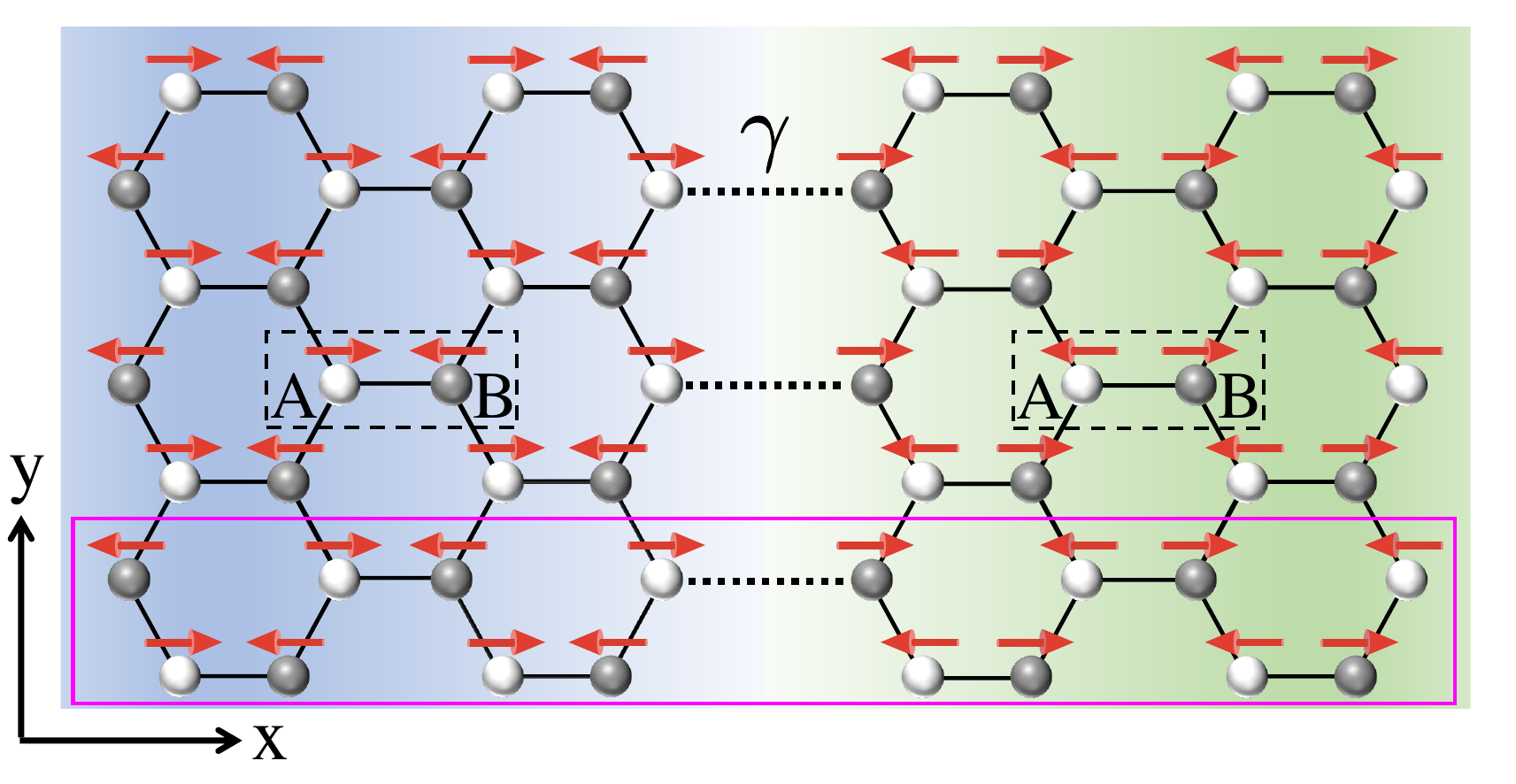}
	\caption{Schematic of a heterojunction composed of two graphene
	nanoflakes with opposite in-plane antiferromagnetic orderings.
	The graphene lattice with unit cell (dashed square) consists of two distinct sublattice sites, $A$ (white ball) and $B$ (grey ball) that prefer to localize opposite magnetic orderings, making the whole system antiferromagnetic with the same number of $A$ and $B$ sublattice sites.
	The magnetic ordering directions of the sublattice sites $A$ and $B$
	in the blue (green) region are respectively along the $+x$($-x$)-axis
	and $-x$($+x$)-axis, as indicated by red arrows.
	Under the calculation of the energy band, it is periodic along
	the $y$ direction with the super unit cell (pink box).
	$\gamma$ is the coupling strength of the two graphene nanoflake regions.}
	\label{fig1}
\end{figure}

We consider a heterojunction composed of two graphene nanoflakes
with opposite in-plane antiferromagnetic orderings, as shown in Fig.~\ref{fig1}.
This graphene heterojunction system can be described by the following
tight-binding Hamiltonian:
\begin{align}
  H=H_{b}+H_{g}+H_{t},
  \label{eq1}
 \end{align}
where $H_{b}$, $H_{g}$, and $H_{t}$
are the Hamiltonian of the blue region, the green region, and
the coupling between them, respectively.
In the tight-binding representation, $H_{b/g}$ can be written by modified Kane-Mele model as follows \cite{Kane2005}
\begin{align}
  H_{b/g}=&\sum_{\left< i,j \right> ,\sigma}t{c_{i\sigma}^{\dag}c_{j\sigma}}+\sum_{\left< \left< i,j \right> \right> ,\sigma ,\sigma^{\prime}}{i t_{s} \nu _{ij} c_{i\sigma}^{\dag}c_{j\sigma^{\prime}}\left[ \mathbf{\hat{s}}_{z} \right] _{\sigma \sigma^{\prime}}}\nonumber \\
  &+\sum_{i,\sigma ,\sigma^{\prime}}{\lambda_{i}^{b/g} c_{i\sigma}^{\dag}c_{i\sigma^{\prime}}\left[ \mathbf{\hat{s}}_{x} \right] _{\sigma \sigma^{\prime}}}.
  \label{eq2}
 \end{align}
 Here $c_{i\sigma}$ and $c_{i\sigma}^{\dag}$ are the annihilation and creation operators for an electron with spin $\sigma$ at the $i$th site, $t$ denotes the nearest-neighbor hopping amplitude. The second term is the intrinsic spin-orbit interaction with an amplitude of $t_{s}$, which is associated with the next nearest neighbor hopping. It depends on clockwise ($\nu _{ij}=-1$) or counterclockwise ($\nu _{ij}=1$) hopping paths from site $j$ to $i$.
 The last term in Eq.~(\ref{eq2}) is the in-plane antiferromagnetic ordering along the $x$ direction. $\lambda_{i}^{b/g}$ is the staggered-sublattice antiferromagnetic ordering magnitude in the blue/green region
 ($\lambda_{B}^{b/g}=-\lambda_{A}^{b/g}$). Moreover, the magnetization directions of the blue region and green region at sublattice site $A$ keep opposite ($\lambda_{A}^{b}=-\lambda_{A}^{g}$). $\mathbf{\hat{s}}_{x,y,z}$ are Pauli matrices denoting spin space.

The coupling Hamiltonian $H_{t}$ between the two regions is described by
  \begin{align}
  H_{t}=&\sum_{\left< i^{\prime},j^{\prime} \right> ,\sigma}\gamma t{c_{bi^{\prime}\sigma}^{\dag}c_{gj^{\prime}\sigma}}\nonumber \\
  &+\sum_{\left< \left< i^{\prime},j^{\prime} \right> \right> ,\sigma ,\sigma^{\prime}}{i \gamma t_{s} \nu _{ij} c_{bi^{\prime}\sigma}^{\dag}c_{gj^{\prime}\sigma^{\prime}}\left[ \mathbf{\hat{s}}_{z} \right] _{\sigma \sigma^{\prime}}}+H.c.,
  \label{eq3}
 \end{align}
 where $c_{bi^{\prime}\sigma}^{\dag}$ ($c_{gi^{\prime}\sigma}$) is the creation (annihilation) operator for an electron with spin $\sigma$ at the $i^{\prime}$th near-interface site of the blue (green) region.
 $\gamma$ is the coupling strength of the two graphene nanoflake regions. Without loss of generality, we set $t=1$ as an energy unit,
 and $t_s=0.1t$, $\lambda_{A}^{b}=-\lambda_{A}^{g}=0.2t,~\lambda_{B}^{b/g}=-\lambda_{A}^{b/g}$, and $\gamma=1$ in our calculations unless otherwise noted.

%========================================================================================
\section{\uppercase\expandafter{\romannumeral 3}.results and discussion}
%========================================================================================
\subsection{\textbf{A. Topologically-protected 0D in-gap interface end states induced by heterojunction}}

\begin{figure}
	\centering
	\includegraphics[width=8.6cm,angle=0]{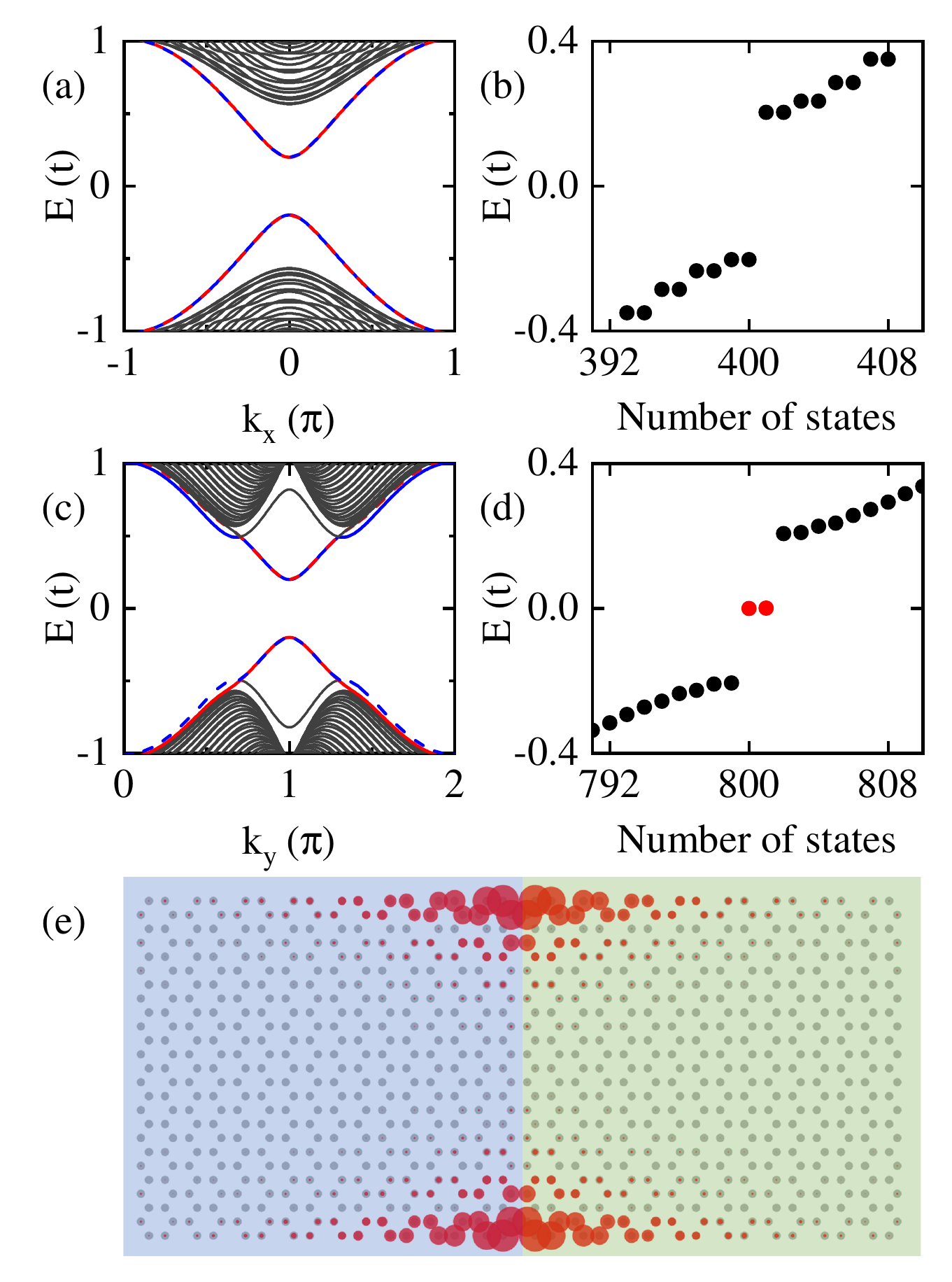}
	\caption{(a) Band spectra for the armchair nanoribbon
with $x$-direction antiferromagnetic ordering.
(b) Energy levels of square-shaped nanoflake with antiferromagnetic ordering.
(c) Band spectra for the heterojunction zigzag nanoribbon with the $x$-direction opposite antiferromagnetic orderings.
(d) Energy levels of square-shaped heterojunction sample
with both $x$ and $y$ directions taking open boundary conditions.
(e) Probability distribution of the in-gap states.  The area of the red circles are proportional to the charge.
The alternating red and blue lines in (a) and (c) represent
the gapped edge states.
The black and red dots in (b) and (d) correspond to
the eigenenergies of the gapped edge states and in-gap states.
Chosen parameters are $t=1,~t_{s}=0.1t,~\lambda_{A}^{b}=0.2t,~\lambda_{A}^{g}=-0.2t,
~\lambda_{B}^{b/g}=-\lambda_{A}^{b/g}$ and $\gamma=1$.
The size parameters are nanoribbon width $N=80$ for (a) and (c),
the nanoflake size $16 \times 25$ for (b),
the nanoflake size $32 \times 25$ for (d) and (e).
The gray dots in (e) represent the carbon atoms
and their number are $32 \times 25$.
}
\label{fig2}
\end{figure}

First, we study the energy band structure of the individual
antiferromagnetic graphene which is described by the Hamiltonian $H_b$ in Eq.~(\ref{eq2}).
Figure~\ref{fig2}(a) shows the energy bands of the graphene
nanoribbon with armchair-type edges and the parameters $\lambda_{A}=0.2t$, $\lambda_{B}=-0.2t$ and nanoribbon width $N=80$.
One can see that the helical gapless edge states are gapped
in the presence of the antiferromagnetic ordering.
The reason is that the time-reversal symmetry and
the additional mirror symmetry along the armchair edges
are broken by the inducing of antiferromagnetic ordering \cite{Zhu2023}.
It has also been shown in Ref.~\cite{Miao2022} that
the helical gapless edge states are also gapped out along zigzag edges
in the presence of antiferromagnetic ordering.
In Fig.~\ref{fig2}(b), we plot the energy levels of
a square-shaped nanoflake with $y$ ($x$)-direction zigzag (armchair) edge,
where the nanoflake size is set to be $16 \times 25$.
It can be seen that a full energy gap occurs near
the zero energy, and there is no states in the gap.
This indicates no in-gap corner states in the individual
antiferromagnetic graphene nanoflake.

Naturally, the heterojunction could form when two graphene nanoflakes
with different antiferromagnetic orderings couple together.
We construct a heterojunction composed of two graphene nanoflakes,
where the magnetic ordering directions are shown in Fig.~\ref{fig1}.
In Fig.~\ref{fig2}(c), we select the super unit cell (pink box)
in Fig.~\ref{fig1} to plot the energy band structure
of the zigzag graphene heterojunction nanoribbon.
One can see that the helical gapless edge states
of heterojunction are gapped by the antiferromagnetic
orderings $\lambda_{A/B}^{b/g}$.
In addition, we calculate the energy levels of the heterojunction composed
of two square-shaped graphene nanoflakes with opposite in-plane
antiferromagnetic orderings, where the finite size is set to be $32 \times 25$.
It is interesting that two zero-energy in-gap states arise,
as displayed by the red dots in Fig.~\ref{fig2}(d).
It is worth emphasizing that there is no in-gap state in each graphene
nanoflakes when two nanoflakes are uncoupled, see Fig.~\ref{fig2}(b).
But when they are coupled, two zero-energy in-gap states appear.
The wave function probability distributions of the in-gap states
at half-filling are highlighted in Fig.~\ref{fig2}(e).
One can see that the zero-energy in-gap states with fractional charge $e/2$ are almost (due to finite system size) localized at both ends of the interface
in the square-shaped heterojunction.
In particular, the zero-energy in-gap states still appear at the ends of the interface even if the antiferromagnetic ordering amplitude changes smoothly across the interface.
This indicates that the in-gap interface end states induced
by heterostructure are the 0D and zero-energy states.
In addition, topologically protected 0D zero-energy in-gap end states can be induced by heterostructure with antiferromagnetic orderings in any in-plane direction. Without loss of generality, we set the antiferromagnetic orderings direction along the $x$-axis in the following calculations.
\begin{figure}
	\centering
	\includegraphics[width=8.6cm,angle=0]{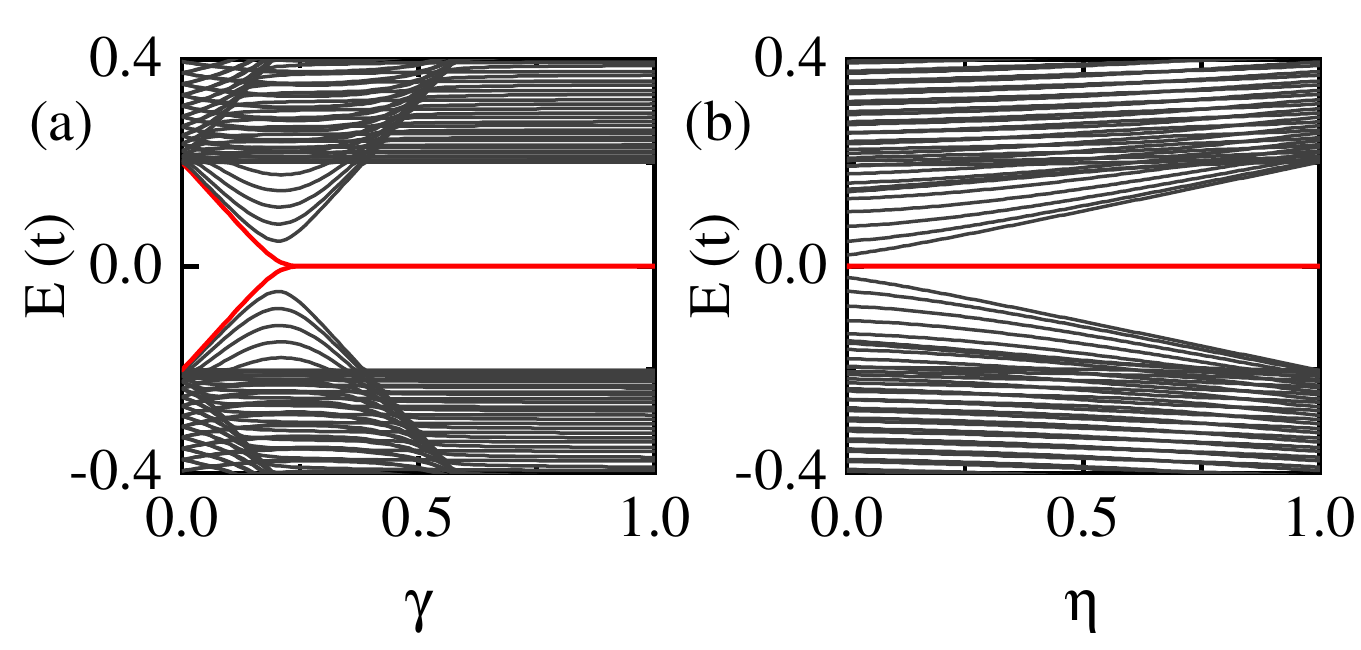}
	\caption{The energy spectrum of square-shaped finite-size heterojunction
nanoflake with opposite antiferromagnetic orderings versus
(a) coupling strength $\gamma$ and (b) ferrimagnetic strength $\eta$ for $\lambda_{B}^{b/g}=-\eta \lambda_{A}^{b/g}$.
The black and red lines indicate the gapped edge states and in-gap states, respectively. The nanoflake size is set to be $40 \times 101$. Other parameters are the same as those in Fig.~\ref{fig2}(d).}
	\label{fig3}
\end{figure}

\begin{figure}
	\centering
	\includegraphics[width=8.6cm,angle=0]{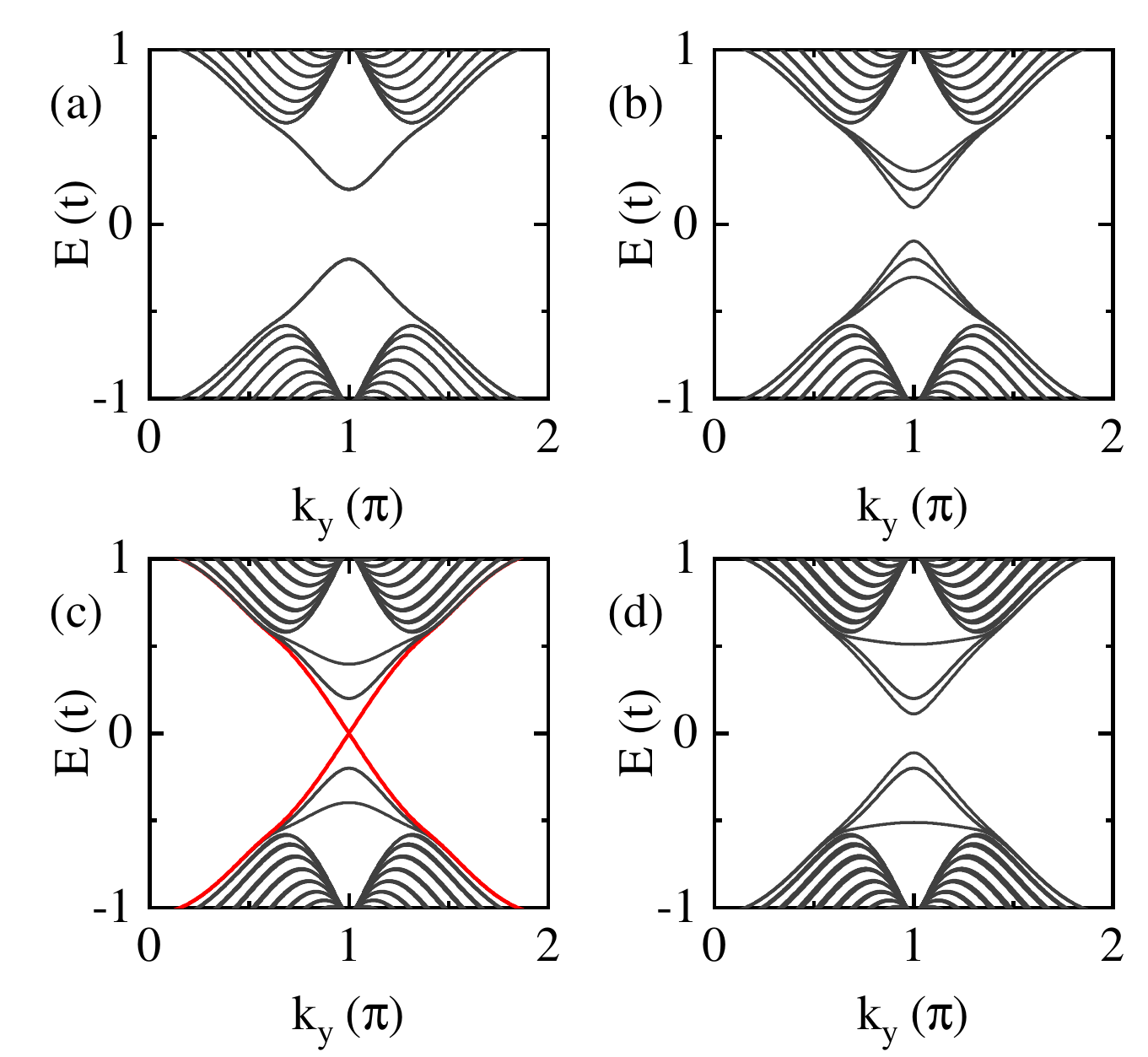}
\caption{Energy band structures for the heterojunction zigzag nanoribbon
with $x$-direction opposite antiferromagnetic orderings.
The coupling strengths are respectively $\gamma=0$ for (a),
$\gamma=0.1$ for (b), $\gamma=0.19$ for (c) and $\gamma=0.3$ for (d).
Other parameters are the same as those in Fig.~\ref{fig2}(c)}
\label{fig4}
\end{figure}

%topologically-protected
To verify that the formation of heterojunction induces
the emergence of 0D in-gap interface end states,
we plot the energy spectrum of the finite-size $32 \times 25$
heterojunction composed of two square-shaped graphene nanoflakes
as a function of coupling strength $\gamma$ in Fig.~\ref{fig3}(a),
and the energy band structure of the heterojunction graphene
nanoribbon for the different coupling strength $\gamma$ in
Figs.~\ref{fig4}(a)-\ref{fig4}(d).
At $\gamma=0$, the two graphene nanoflakes (nanoribbons) building the heterojunction system are completely separate.
In this case, the gapless edge states of the nanoribbon system
could be gapped by the antiferromagnetic orderings
as shown in Fig.~\ref{fig4}(a),
and for the nanoflake system there is no states in the edge gap,
see Fig.~\ref{fig2}(b) or Fig.~\ref{fig3}(a) at $\gamma=0$.
As the coupling strength $\gamma$ increases,
the energy gap of edge states gradually decreases,
as shown by the red lines in Fig.~\ref{fig3}(a)
or Fig.~\ref{fig4}(b).
When $\gamma=0.19$, energy gap of edge states is completely closed
[Fig.~\ref{fig4}(c)], and the gap reopens
with the further increase of the coupling strength $\gamma$
[see Fig.~\ref{fig4}(d) with $\gamma=0.3$ and
Fig.~\ref{fig2}(d) with $\gamma=1$].
In particular, as the energy gap of the edge states reopens
at the large $\gamma$,
two in-gap zero energy states appear
[see the red lines in Fig.~\ref{fig3}(a) and
the red dots in Fig.~\ref{fig2}(d)].
These results can clearly illustrate that
the 0D in-gap interface end states appear in the
coupling graphene heterojunction nanoflake and
they should be topologically protected
from the close and reopen of the energy gap.

When the antiferromagnetic ordering magnitudes $\lambda_{A/B}^{b/g}$
reduce from $\pm 0.2t$ to $0$,
the gap of the gapped edge states gradually decreases and
the in-gap interface end states gradually diffuse
over a relatively large range.
While $\lambda_{A/B}^{b/g}$ reaches zero, the edge states recover gapless
and the interface end states smoothly evolve into
the continuous helical edge states.

Actually, the magnetic orderings of sublattices $A$ and $B$ are
opposite in direction and maybe different in magnitude,
which has a net ferrimagnetic ordering
and can be induced by spontaneous magnetization
in graphene \cite{Son2006, Jung2009}.
Based on this, we set $\lambda_{B}^{b/g}=-\eta \lambda_{A}^{b/g}$.
Figure~\ref{fig3}(b) shows the energy levels of the heterojunction
graphene nanoflake as a function of $\eta$.
Here the coupling strength $\gamma=1$ and
other parameters are the same as Fig.~\ref{fig3}(a).
One can see that the topologically-protected 0D in-gap interface
end states (red lines) remain stable in the process of $\eta$ change.
With the decrease of $\eta$ from $1$ to $0$,
the energy gap of the edge states gradually becomes smaller,
but the in-gap states hold always.
Until $\eta=0$, the energy gap closes and then the in-gap states disappear.
Thus, our results validate topologically-protected
0D in-gap interface end states can be induced by the formation of
heterojunction as long as the magnetic orderings of sublattices $A$ and $B$
are in opposite directions, regardless of whether the magnitudes
are equal or not.

\begin{figure}
	\centering
	\includegraphics[width=8.6cm,angle=0]{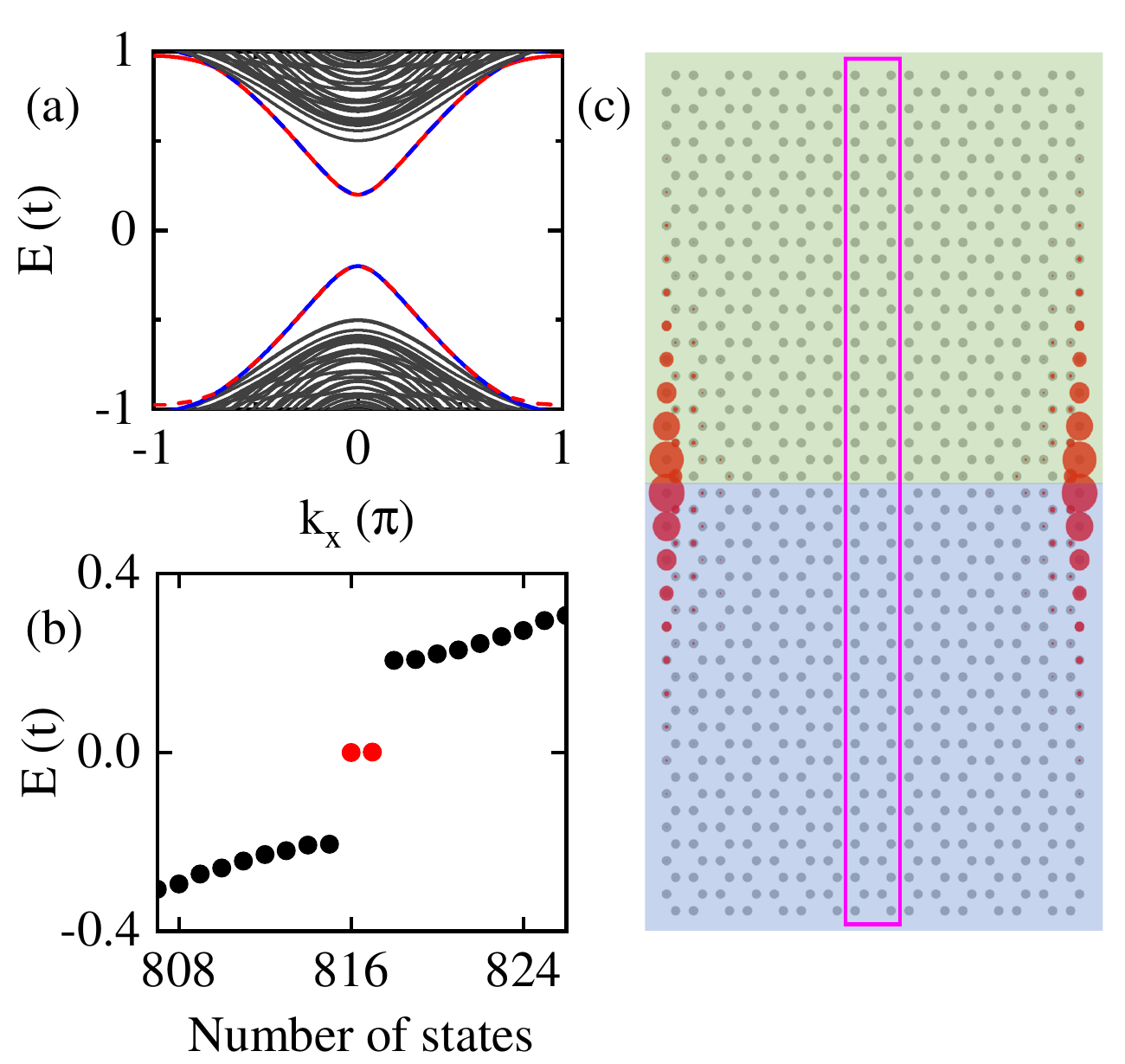}
	\caption{(a) Band spectra for the heterojunction armchair nanoribbon with $x$-direction opposite antiferromagnetic orderings. It is periodic along the $x$ direction with the super unit cell [pink box in (c)]. The alternating red and blue lines represent the gapped edge states.
(b) Energy levels of square-shaped heterojunction nanoflake with
the coupling along the armchair edge.
The black and red dots correspond to the eigenenergies
of the gapped edge states and in-gap states.
(c) Probability distribution of the in-gap states. The area of the red circles are proportional to the charge.
The size parameters are nanoribbon width $N=80$ for (a),
and the nanoflake size $51 \times 16$ for (b) and (c).
The other parameters are the same as those in Fig.~\ref{fig2}(c).}
	\label{fig5}
\end{figure}

Furthermore, we construct another type of heterojunction composed
of two square-shaped graphene nanoflakes
with their interface along the armchair edge, as shown in Fig.~\ref{fig5}(c).
Here the magnetic ordering directions of the sublattice
sites $A$ and $B$ in the blue (green) region are along
the $+x$($-x$)-axis and $-x$($+x$)-axis still.
Figure~\ref{fig5}(a) shows the energy band structure of
the armchair heterojunction graphene nanoribbon.
One can see that the helical gapless edge states
are gapped out along armchair edges also.
In addition, we plot the energy levels of square-shaped finite-size
$51 \times 16$ heterojunction nanoflake in Fig.~\ref{fig5}(b).
The topologically-protected zero-energy in-gap states exist,
as displayed by the red dots.
The wave function probability distribution of the zero-energy
in-gap states at half-filling is highlighted in Fig.~\ref{fig5}(c),
they are still localized at both ends of the interface in the heterojunction.
It is further indicated that topologically-protected 0D in-gap
interface end states can be induced by heterostructure
independent of the interface edge type.

\subsection{\textbf{B. Effects of magnetic disorder, Anderson disorder and interfacial magnetic defects on the in-gap interface end states}}

\begin{figure}
	\centering
	\includegraphics[width=8.6cm,angle=0]{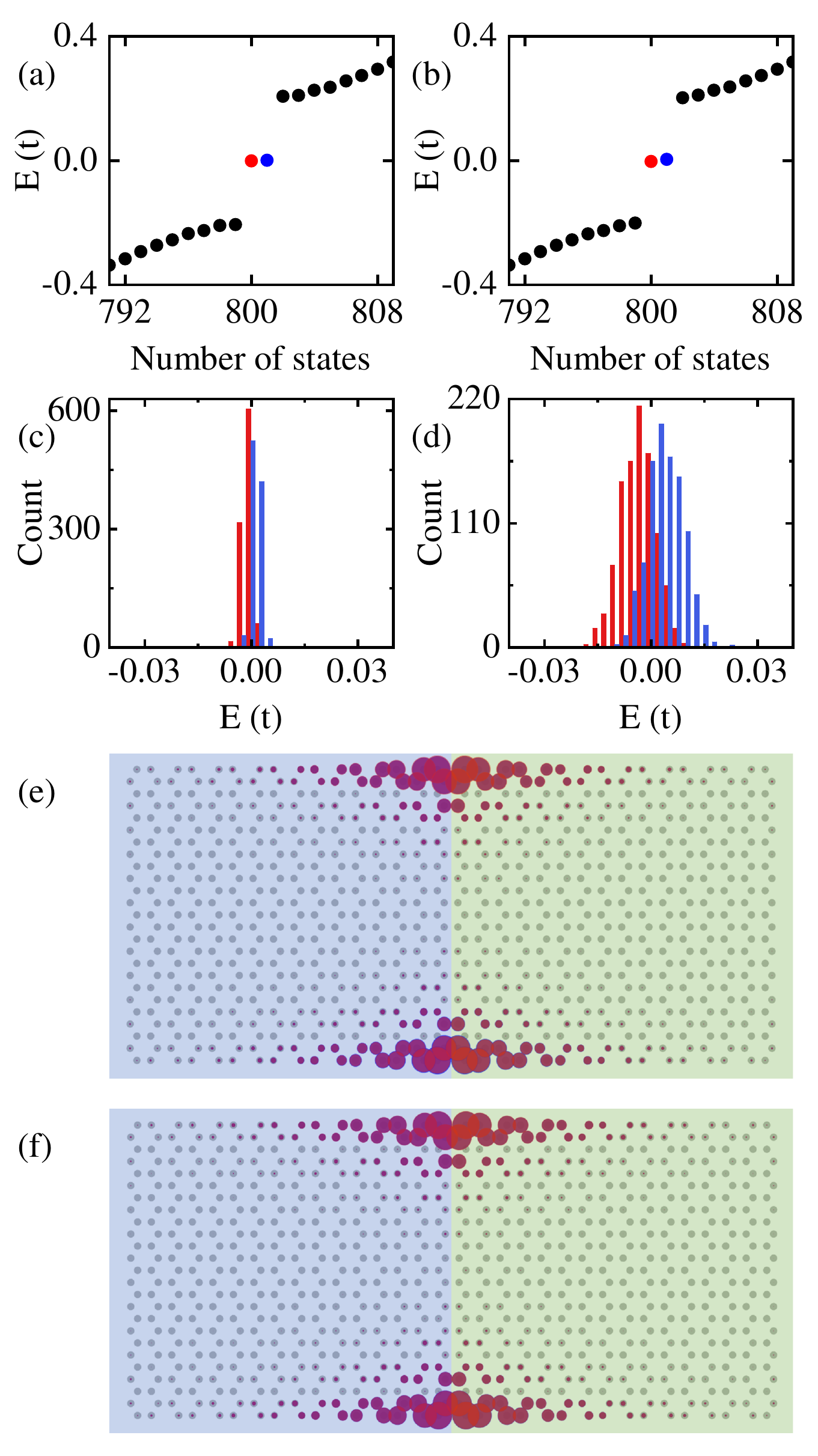}
	\caption{ (a) and (b) show
the energy levels of square-shaped finite-size heterojunction
with different magnetic disorder strength $W_{m}=0.1$ for (a) and $W_{m}=0.4$ for (b). The in-gap states are highlighted in red and blue.
(c) and (d) are the energy distribution of two in-gap states.
(e) and (f) show the wave function distributions of in-gap states under average calculation. 
%{\color{blue} The area of the red and blue circles are proportional to the charge.}
The magnetic disorder strength $W_{m}=0.1$ for (a), (c) and (e)
and $W_{m}=0.4$ for (b), (d) and (f).
Other parameters are the same as those in Fig.~\ref{fig2}(d,e).}
	\label{fig6}
\end{figure}

In this subsection, we study the robustness of the in-gap interface end states.
The effects of various disorders, including magnetic disorder, Anderson disorder
and interfacial magnetic defects, on the in-gap states are investigated.

It is shown that the magnitude of edge antiferromagnetic induced
by spontaneous magnetization is inhomogeneous by several works
applying the mean field theory calculations and
quantum Monte Carlo simulations \cite{Weymann2016,Krompiewski2017,Raczkowski2020,Phung2020,Phung2022}.
In Fig.~\ref{fig6}, we calculate the energy levels
and probability distribution of the square-shaped heterojunction system
with different magnetic disorder strength $W_{m}$.
In the presence of the magnetic disorder, the disorder term
$\sum_{i,\sigma ,\sigma^{\prime}}{\omega \lambda_{i}^{b/g} c_{i\sigma}^{\dag}c_{i\sigma^{\prime}}\left[ \mathbf{\hat{m}^{\prime}}\cdot \mathbf{\hat{s}} \right] _{\sigma \sigma^{\prime}}}$ is added to Hamiltonian in Eq.~(\ref{eq2}),
where $\omega$ is randomly distributed in the interval
$\left[0, W_{m}\right]$ and the unit vector
$\mathbf{\hat{m}^{\prime}} =\left(\cos \theta^{\prime}, \sin \theta^{\prime}\right)$
with angle $\theta^{\prime}$ being randomly distributed in the interval $\left[0, 2\pi\right]$.
Thus, the numerical magnitude of the magnetic disorder is distributed in the interval $\left[0, W_{m}\lambda_{i}^{b/g}\right]$, and the direction of the magnetic disorder is arbitrarily random in the plane.
All the curves are averaged over $1000$ random configurations,
which is enough to obtain reasonable results.
From Fig.~\ref{fig6}(a), one can see that the topologically-protected
0D in-gap interface end states still emerge at the magnetic disorder strength $W_{m}=0.1$, which are highlighted by red and blue dots.
At $W_{m}=0.4$, the energy degeneracy of the in-gap interface end states
is slightly broken in Fig.~\ref{fig6}(b).
Figures ~\ref{fig6}(c) and ~\ref{fig6}(d) show that
the energy value distribution of the two in-gap interface end states
with different magnetic disorder strength $W_{m}$.
The red and blue histograms of energy distribution
in Figs.~\ref{fig6}(c) and ~\ref{fig6}(d) correspond to
the in-gap states highlighted in red and blue [see Figs.~\ref{fig6}(a) and ~\ref{fig6}(b)], respectively.
In Fig.~\ref{fig6}(c), one can see that the energy values
are clustered around zero with $W_{m}=0.1$.
At $W_{m}=0.4$, the energy values of the two in-gap states
are slightly diffused [see Fig.~\ref{fig6}(d)],
but they still distribute at a small range $\left(-0.02t, 0.02t\right)$
which is much smaller than the edge gap $\left(-0.15t, 0.15t\right)$.
The wave function distributions of in-gap states under average calculation
are shown in Figs.~\ref{fig6}(e) and ~\ref{fig6}(f)
with different magnetic disorder strength $W_{m}$.
Whether $W_{m}=0.1$ or $W_{m}=0.4$, the in-gap states are uniformly
distributed at two ends of the interface of heterojunction.
The above results show that moderate magnetic fluctuation
does not suppress the appearance of topologically-protected 0D
in-gap interface end states, but could slightly lift their degeneracy.
It can be concluded that the topologically-protected 0D in-gap interface end
state is robust against the magnetic disorder.

\begin{figure}
	\centering
	\includegraphics[width=8.6cm,angle=0]{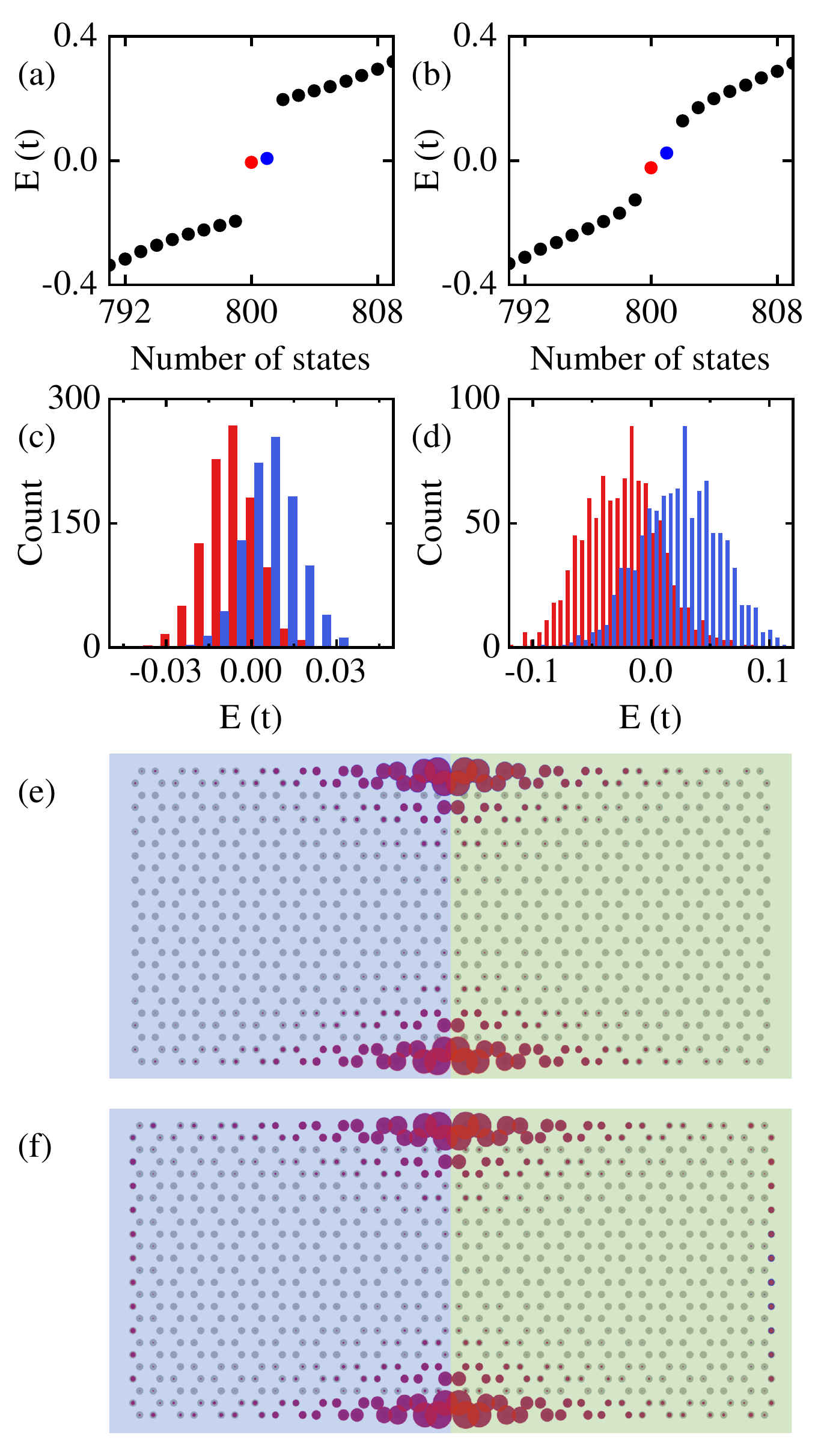}
	\caption{(a) and (b) show the energy levels of square-shaped finite-size heterojunction nanoflakes with different Anderson disorder strength $W_{a}=0.1$ for (a) and $W_{a}=0.4$ for (b). The in-gap states are highlighted in red and blue. (c) and (d) are the energy distribution of two in-gap states.
(e) and (f) show the wave function distributions of in-gap states under average calculation.
%{\color{blue} The area of the red and blue circles are proportional to the charge.}
The Anderson disorder strength $W_{a}=0.1$ for (a), (c) and (e)
and $W_{a}=0.4$ for (b), (d) and (f).
Other parameters are the same as those in Fig.~\ref{fig2}(d,e).}
	\label{fig7}
\end{figure}

 \begin{figure}
%	\centering
	\includegraphics[width=8.6cm,clip]{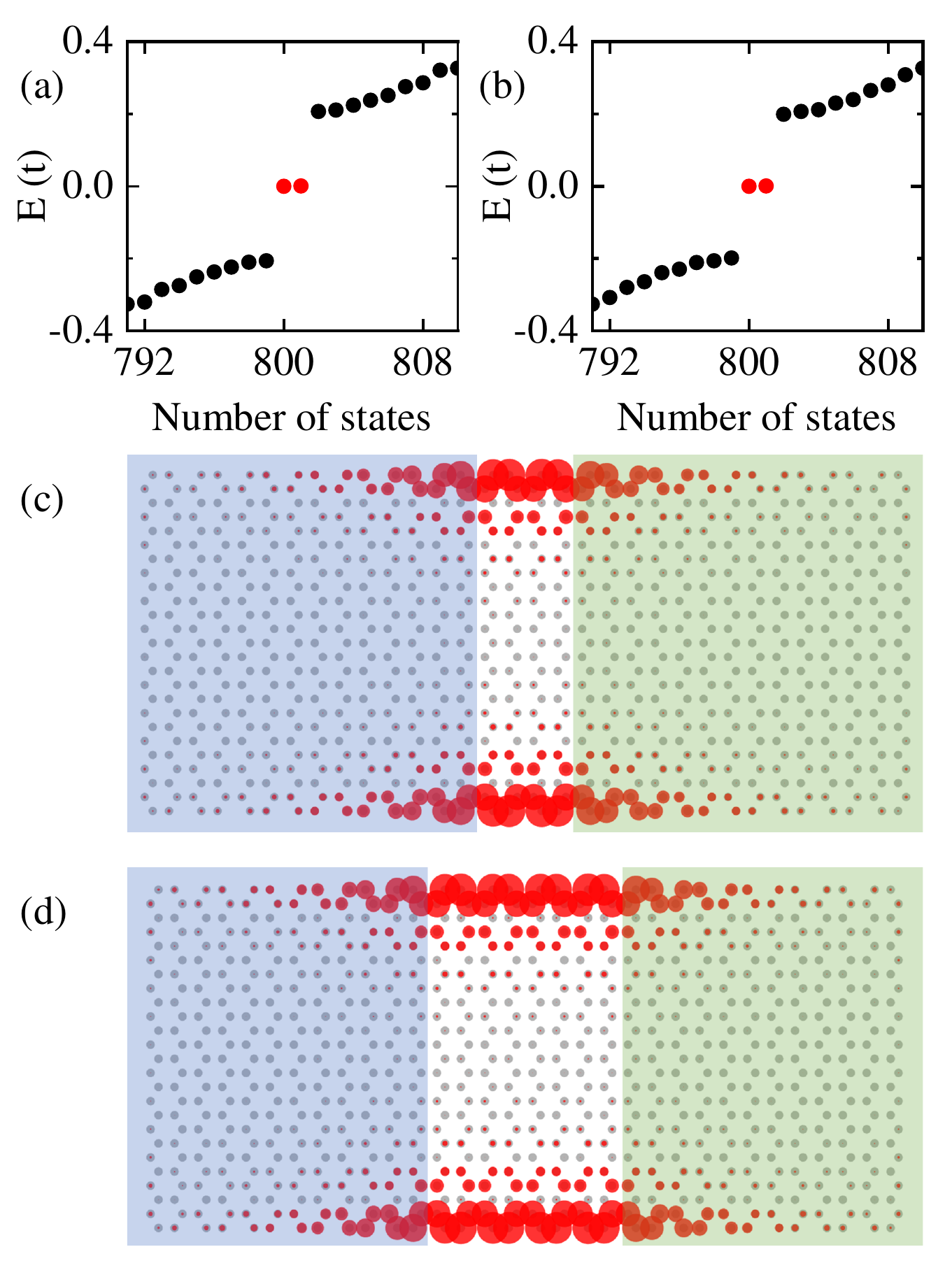}
	\caption{Energy levels and wave function distribution of
square-shaped finite-size heterojunction nanoflakes
with different interfacial magnetic defects width $D=4$ for (a) and (c)
and $D=8$ for (b) and (d).
The in-gap states are highlighted in red. 
%{\color{blue} The area of the red circles in (c) and (d) are proportional to the charge.}
The magnetic ordering directions of the sublattice sites $A$ and $B$
in the blue (green) region are along the $+x$($-x$)-axis and $-x$($+x$)-axis,
respectively. The lattice sites in the white region are not magnetic.
Other parameters are the same as those in Fig.~\ref{fig2}(d,e).}
	\label{fig8}
\end{figure}

\begin{figure*}
	\centering
	\includegraphics[width=2\columnwidth,clip]{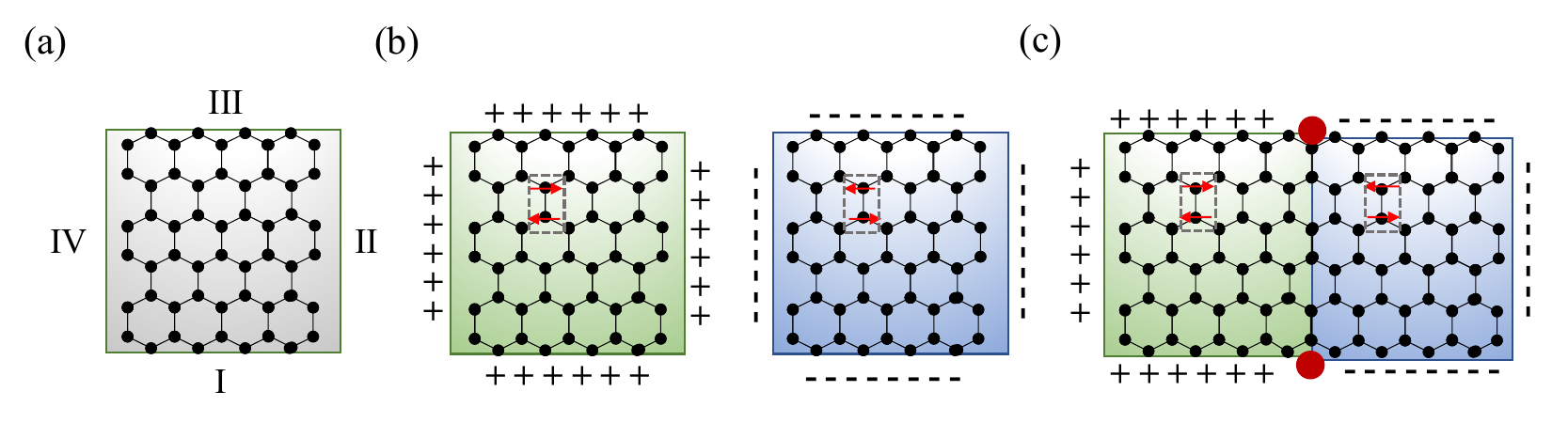}
	\caption{(a) Illustration of rectangular graphene, with $I-IV$ denoting the four boundaries. (b) Graphene in the blue and green regions exhibits opposite in-plane antiferromagnetic orderings. The $+ / -$ signs at the boundaries represent the signs of the effective mass terms.
(c) 0D in-gap interface end states (indicated by red circles)
emerge at the attachment points of the blue and green regions.
}
	\label{fig9}
\end{figure*}

To verify the robustness of the topologically-protected 0D in-gap interface end states, we also calculate the energy levels and probability distribution of the heterojunction system with different Anderson disorder strength $W_{a}$ in Fig.~\ref{fig7}.
The Anderson disorder term  $\sum_{i,\sigma}{\omega_{i} c_{i\sigma}^{\dag}c_{i\sigma}}$ is added to Hamiltonian in Eq.~(\ref{eq2})
with $\omega_{i}$ being uniformly distributed in the interval $\left[-W_{a}, W_{a}\right]$.
All the curves in Fig.~\ref{fig7} are averaged over $1000$ random configurations, which is enough to obtain reasonable results.
It is seen that there are two in-gap states with non-degeneracy and non-zero-energy values, the degeneracy of in-gap states is slightly broken
at the Anderson disorder strength $W_{a}=0.1$ in Fig.~\ref{fig7}(a), which are highlighted by red and blue dots.
There still are two in-gap states with non-degeneracy and non-zero-energy values
and the energy values of the gapped edge states (black dots) tend to the zero energy at $W_{a}=0.4$ in Fig.~\ref{fig7}(b).
Figures ~\ref{fig7}(c) and ~\ref{fig7}(d) show the energy value distribution
of the two in-gap states with different Anderson disorder strength $W_{a}$.
The red and blue histograms of energy distribution in Figs.~\ref{fig7}(c) and ~\ref{fig7}(d) correspond to the in-gap states highlighted in red and blue [see Figs.~\ref{fig7}(a) and ~\ref{fig7}(b)], respectively.
In Fig.~\ref{fig7}(c), one can see that the energy values are
clustered around zero with $W_{a}=0.1$.
The energy values of the two in-gap states are highly diffused
with $W_{a}=0.4$ [see Fig.~\ref{fig7}(d)], but they are still smaller than the edge gap.
The wave function distributions of in-gap states under average calculation are shown
in Figs.~\ref{fig7}(e) and ~\ref{fig7}(f) with
different Anderson disorder strength $W_{a}$.
Whether $W_{a}=0.1$ or $W_{a}=0.4$, the in-gap states are perfectly
distributed at two ends of the interface of heterojunction.
It can be concluded that the topologically-protected
0D in-gap interface end state is robust against the Anderson disorder.

Interfacial magnetic defects are also a common perturbation,
so we consider the atoms with an intermediate interface width of $D$ without magnetism.
We plot the energy levels of square-shaped heterojunction nanoflakes
with different magnetism defects covering the intermediate interface width
of $D$ in Fig.~\ref{fig8}.
Figures~\ref{fig8}(a) and ~\ref{fig8}(b) show that the red zero-energy
degenerate in-gap states are both stable at $D=4$ and $D=8$.
As is shown in Fig.~\ref{fig8}(c), the zero-energy in-gap states
at the interface diffuse from magnetic lattices in a certain range
in the $x$-direction, and there are also localized at non-magnetic lattices with $D=4$.
The bound states have a larger transverse diffusion range for $D=8$,
and the lattices at two ends of the intermediate junction
with the zero magnetism have the large wave function
distribution [see Fig.~\ref{fig8}(d)].
This indicates that the topologically-protected 0D in-gap interface end states
induced by heterojunction coupling are robust against interfacial magnetic defects.

\subsection{\textbf{C. The origin of topologically-protected 0D in-gap interface end states}}

To gain an intuitive understanding of the topologically-protected
0D in-gap interface end states, we study the helical edge states in the
antiferromagnetic graphene system.
When the system is a Kane-Mele model
in the absence of the ferromagnetic orderings,
boundary states at zigzag and armchair boundaries can be resolved analytically \cite{doh2014analytic}.
We label the four edges of a square as $I $, $ {II} $, $ {III} $, and $ {IV} $
in Fig.~\ref{fig9}(a), and the corresponding spin-helical edge states (spinor part) are given by \cite{tanTwoparticleBerryPhase2022}
\begin{eqnarray}
{\left| {{\chi _ \uparrow }} \right\rangle ^{(I)}} &=
&\frac{1}{{\sqrt {1 + \mu } }}{\left[ {\begin{array}{*{20}{c}}
1\\
{ - i\sqrt \mu  }
\end{array}} \right]_\sigma } \otimes {\left[ {\begin{array}{*{20}{l}}
1\\
0
\end{array}} \right]_s} \nonumber\\
{\left| {{\chi _ \downarrow }} \right\rangle ^{(I)}} &= &\frac{1}{{\sqrt {1 + \mu } }}{\left[ {\begin{array}{*{20}{c}}
1\\
{i\sqrt \mu  }
\end{array}} \right]_\sigma } \otimes {\left[ {\begin{array}{*{20}{l}}
0\\
1
\end{array}} \right]_s} \nonumber\\
{\left| {{\chi _ \uparrow }} \right\rangle ^{(II)}} &=&
{\left[ {\begin{array}{*{20}{c}}
1\\
{ - i}
\end{array}} \right]_\sigma } \otimes {\left[ {\begin{array}{*{20}{l}}
1\\
0
\end{array}} \right]_s},~~ {\left| {{\chi _ \downarrow }} \right\rangle ^{(II)}} = {\left[ {\begin{array}{*{20}{l}}
1\\
i
\end{array}} \right]_\sigma } \otimes {\left[ {\begin{array}{*{20}{l}}
0\\
1
\end{array}} \right]_s}\nonumber\\
{\left| {{\chi _ \uparrow }} \right\rangle ^{(III)}}& =
&\frac{1}{{\sqrt {1 + 1/\mu } }}{\left[ {\begin{array}{*{20}{c}}
1\\
{ - \frac{i}{{\sqrt \mu  }}}
\end{array}} \right]_\sigma } \otimes {\left[ {\begin{array}{*{20}{l}}
1\\
0
\end{array}} \right]_s} \nonumber\\
{\left| {{\chi _ \downarrow }} \right\rangle ^{(III)}} &=
&\frac{1}{{\sqrt {1 + 1/\mu } }}{\left[ {\begin{array}{*{20}{c}}
1\\
{\frac{i}{{\sqrt \mu  }}}
\end{array}} \right]_\sigma } \otimes {\left[ {\begin{array}{*{20}{l}}
0\\
1
\end{array}} \right]_s}\nonumber\\
{\left| {{\chi _ \uparrow }} \right\rangle ^{(IV)}} &=&
{\left[ {\begin{array}{*{20}{c}}
1\\
i
\end{array}} \right]_\sigma } \otimes {\left[ {\begin{array}{*{20}{l}}
1\\
0
\end{array}} \right]_s},~~{\left| {{\chi _ \downarrow }} \right\rangle ^{(IV)}} = {\left[ {\begin{array}{*{20}{l}}
1\\
{ - i}
\end{array}} \right]_\sigma } \otimes {\left[ {\begin{array}{*{20}{l}}
0\\
1
\end{array}} \right]_s}\nonumber
\end{eqnarray}
with $\mu  = \left( {1 + \frac{{{t^2}}}{{8t_s^2}}} \right) - \sqrt {{{\left( {1 + \frac{{{t^2}}}{{8t_s^2}}} \right)}^2} - 1}$,  $\sigma$ and $s$ repersent sublattice and spin.
We focus on the edge $I$ first. The effective mass term ${\bf M}_{I}^{}$ of the edge $I$ can be obtained by projecting the in-plane antiferromagnetic ordering term onto the subspace spanned by $\left| {\chi _ \uparrow ^{\left( I \right)}} \right\rangle$ and $\left| {\chi _ \downarrow ^{\left( I \right)}} \right\rangle$.
\begin{eqnarray}
{{\bf M}_I} &= & {\lambda^{b/g}_A}\left( {\begin{array}{*{20}{c}}
{\langle \chi _ \uparrow ^{\left( I \right)}|{\sigma _z}{s_x}\left| {\chi _ \uparrow ^{\left( I \right)}} \right\rangle }&{\langle \chi _ \uparrow ^{\left( I \right)}|{\sigma _z}{s_x}\left| {\chi _ \downarrow ^{\left( I \right)}} \right\rangle }\\
{\langle \chi _ \downarrow ^{\left( I \right)}|{\sigma _z}{s_x}\left| {\chi _ \uparrow ^{\left( I \right)}} \right\rangle }&{\langle \chi _ \downarrow ^{\left( I \right)}|{\sigma _z}{s_x}\left| {\chi _ \downarrow ^{\left( I \right)}} \right\rangle }
\end{array}} \right)\nonumber\\
& = & {\lambda^{b/g}_A}\frac{{1 + {\rm{|}}\mu {\rm{|}}}}{{{\rm{|1 + }}\mu {\rm{|}}}}\left( {\begin{array}{*{20}{c}}
0&1\\
1&0
\end{array}} \right)  = {\lambda^{b/g}_A}\left( {\begin{array}{*{20}{c}}
0&1\\
1&0
\end{array}} \right)\nonumber\\
& = & m_I^{}\left( {\begin{array}{*{20}{c}}
0&1\\
1&0
\end{array}} \right)
\end{eqnarray}
where $\sigma _i^{}$ and $s_i^{}$ are Pauli matrices in the sublattice
$\left( {A, B} \right)$ and spin $\left( { \uparrow ; \downarrow } \right)$.
So we obtain the effective mass for the edge $I$, $m_I=\lambda^{b/g}_A$.
Similarly, the effective mass term for the remaining three edges can be obtained as $m_{II}^{} = 2{\lambda^{b/g}_A}$, $m_{III}^{} = {\lambda^{b/g}_A}$,
and $m_{IV}^{} = 2{\lambda^{b/g}_A}$.
Due to the mass term for four edges have the same sign,
it follows that gaps will open along all four edges of both the blue region
and the green region, and no in-gap corner states exist.
On the other hand, the effective mass terms in the blue and green regions
have the opposite signs because they have the opposite antiferromagnetic
orderings [see Fig.~\ref{fig9}(b)].
Therefore if the blue region and the green region are attached,
a mass domain wall is obtained on the edge,
resulting in the emergence of topologically-protected 0D in-gap states
at the ends of the interface [see Fig.~\ref{fig9}(c)].

\subsection{\textbf{D. Tunability of topologically-protected 0D in-gap interface end states}}

\begin{figure}
	\centering
	\includegraphics[width=8.6cm,clip]{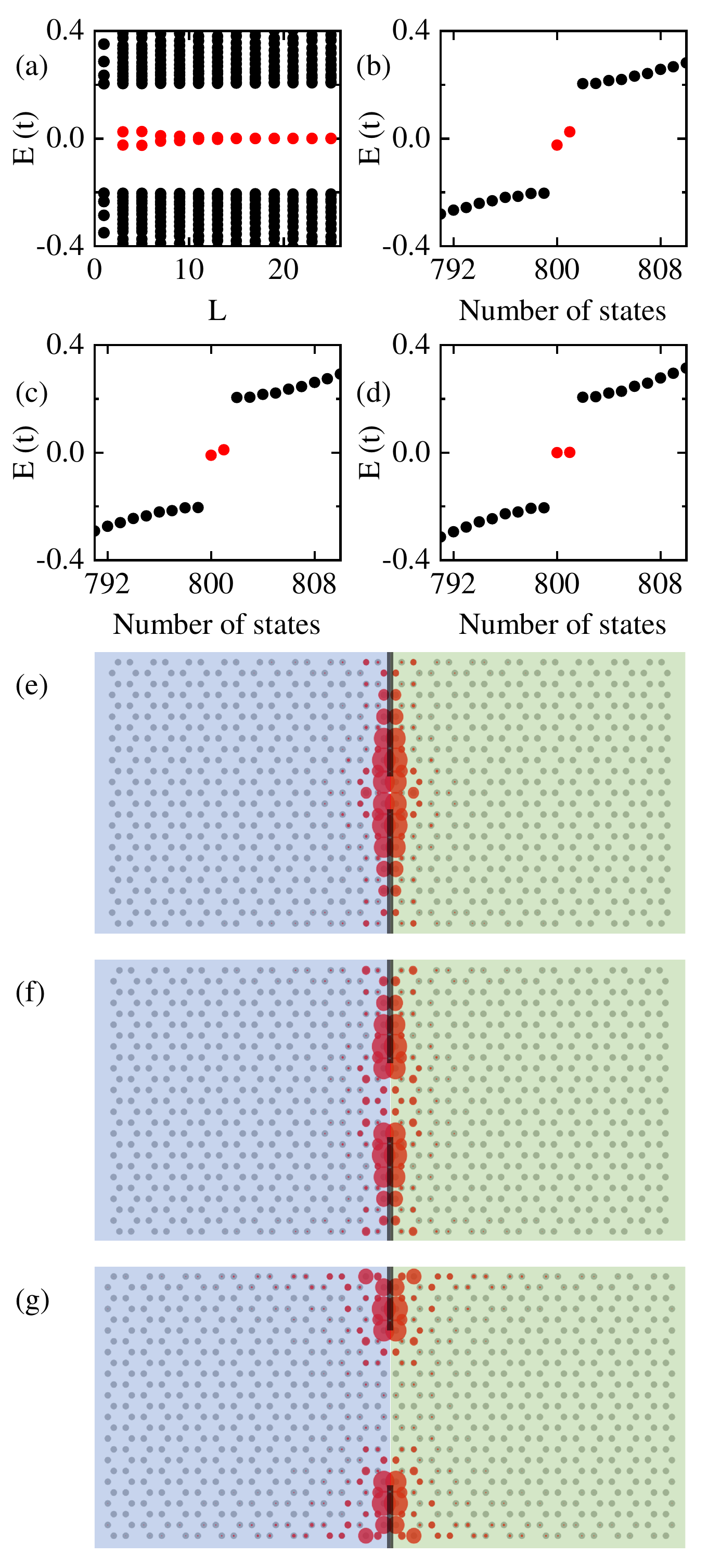}
	\caption{(a) Energy levels of square-shaped finite-size heterojunction nanoflakes as a function of coupling length $L$ along $y$-axis.
The coupling strength $\gamma$ equal to $1$ within the coupling length $L$, otherwise $\gamma=0$. (b-g) Energy levels and wave function distributions of
the in-gap states for the heterojunction nanoflakes with different coupling lengths $L=3$ for (b) (e), $L=7$ for (c) (f), and $L=15$ for (d) (g).
The black lines in (e), (f) and (g) represent the uncoupled interface region. %{\color{blue} The area of the red circles in (e-g) are proportional to the charge.}
The parameters are the same as those in Fig.~\ref{fig2}(d,e).}
	\label{fig10}
\end{figure}

The coupling length and strength between two graphene nanoflakes can be tuned by stain and substrate experimentally \cite{Cheng2023}. In addition, the use of laser scribing and STM to cut off at different coupling positions of the heterojunction is able to achieve 0D interface end states.
In Fig.~\ref{fig10}(a), we plot the energy levels $E$ of the square-shaped finite-size heterojunction nanoflake as a function of coupling length $L$.
Within the coupling width of the interface, the coupling strength is $\gamma=1$, otherwise $\gamma=0$.
The other parameters are the same as Fig.~\ref{fig2}(d).
We plot energy levels and wave function distribution of in-gap states
with different coupling length $L=3$ for (b) (e), $L=7$ for (c) (f) and $L=15$ for (d) (g).
At $L=1$, the two graphene nanoflakes building the heterojunction system
are almost separate, and the energy gap exists as shown in Fig.~\ref{fig10}(a).
At $L=3$, two nonzero-energy in-gap states (red dots) appear [see Fig.~\ref{fig10}(b)] and interact in the middle of the heterojunction interface [see Fig.~\ref{fig10}(e)].
In this case, two in-gap states are partially overlapping in space.
As $L$ increases, the energy value of the in-gap states gradually
approach the zero energy Fermi level.
At $L=7$, there is still a small energy difference between the two in-gap states, as shown in Fig.~\ref{fig10}(c).
Their wave function distributions have decoupled in space to form topologically-protected 0D in-gap interface end states [see Fig.~\ref{fig10}(f)].
Until $L=15$, two topologically-protected 0D in-gap interface end states
with zero-energy degenerate are induced in the heterojunction,
as shown in Fig.~\ref{fig10}(d).
Figure~\ref{fig10}(g) shows the topologically-protected 0D in-gap interface states in space are well localized near the end of the lattices in the heterojunction coupling region.
The above results and analysis indicate that the small $L=3$ is sufficient to induce the in-gap interface states.
When two topologically-protected in-gap interface states are very close to each other, they will not disappear but diffuse within a certain range.
The larger the coupling region is, the energy values of the in-gap states tend to zero energy degenerate. Interestingly, the position of the topologically-protected 0D in-gap interface end states is tunable by tuning
the coupling length.

\begin{figure}
	\centering
	\includegraphics[width=8.6cm,angle=0]{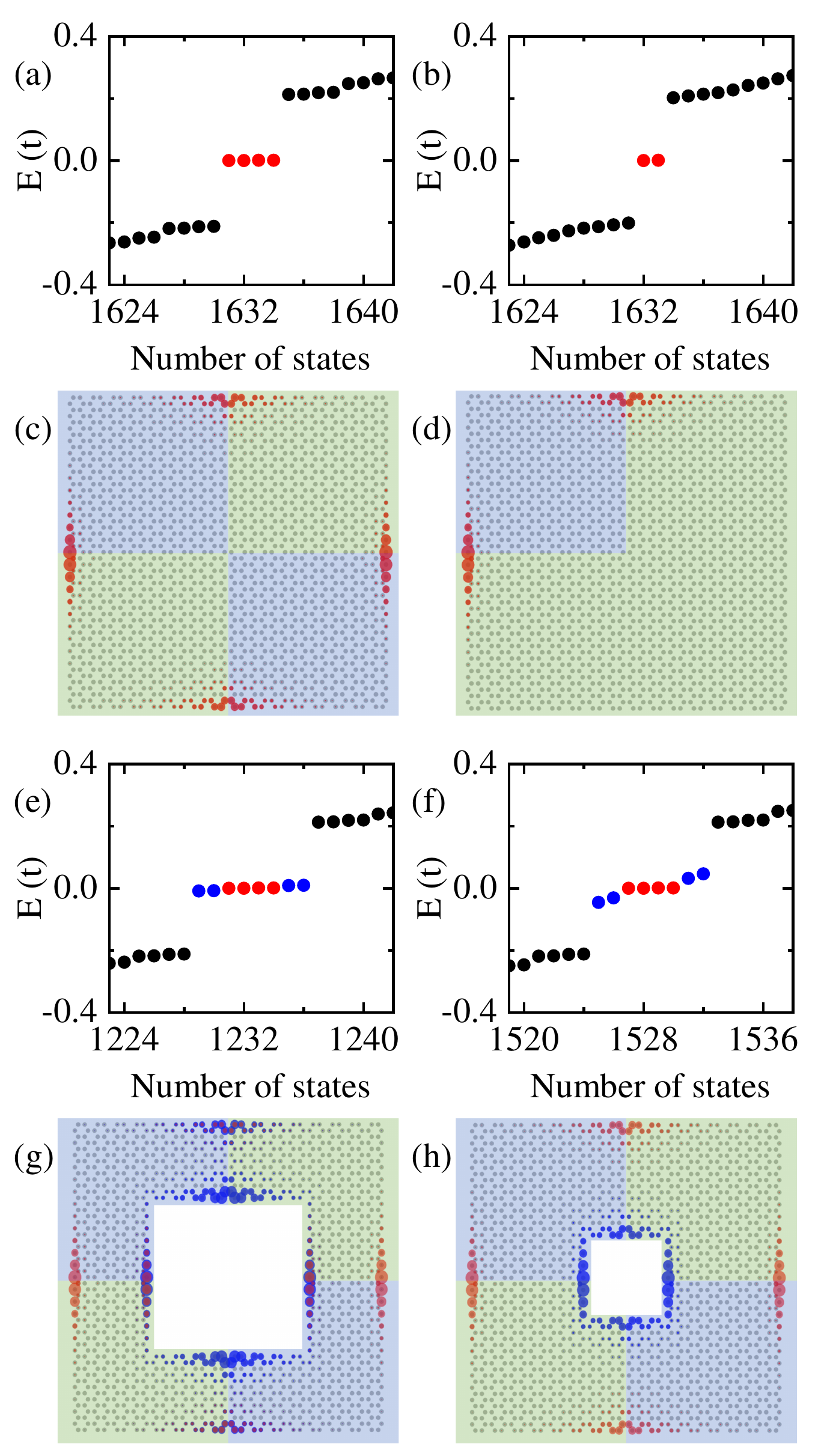}
	\caption{Energy levels and wave function distribution of square-shaped finite-size $32 \times 51$ heterojunction nanoflake consisting of four regions with different antiferromagnetic fields and central hollows.
The parameters are $t=1$, $t_{s}=0.1t$, $\gamma=1$, and
the antiferromagnetic ordering magnitude $\lambda_{A}^{b} =-\lambda_{B}^{b}=0.2t$
in the blue region and $\lambda_{A}^{g} =-\lambda_{B}^{g}=-0.2t$
in the green region. There are no central hollows in (c) and (d). The size of the central hollows is $16 \times 25$ for (g) and $8 \times 13$ for (h). The parameters in (a), (b), (e), and (f) are the same as those in (c), (d), (g), and (h).
The in-gap states are highlighted in red and blue. 
%{\color{blue} The area of the red and blue circles are proportional to the charge.}
}
	\label{fig11}
\end{figure}

In order to adjust the number of topologically-protected 0D in-gap interface end states, we build a heterojunction with four different antiferromagnetic orderings regions, and the direction of the antiferromagnetic orderings can be adjusted separately in Fig.~\ref{fig11}.
For convenience, the top left, top right, bottom left, and bottom right regions are labeled in turn as region-1, region-2, region-3, and region-4 with staggered-sublattice antiferromagnatic ordering $\lambda_{i}^{1},~\lambda_{i}^{2},~\lambda_{i}^{3}$ and $\lambda_{i}^{4}$.
We calculate the energy levels of heterojunction nanoflake consisting of four regions in Figs.~\ref{fig11}(a),~\ref{fig11}(b),~\ref{fig11}(e), and~\ref{fig11}(f) with different magnetic orderings and hollows.
The corresponding density distributions of in-gap states
are shown in Figs.~\ref{fig11}(c),~\ref{fig11}(d),~\ref{fig11}(g), and~\ref{fig11}(h).
The main parameters difference are $\lambda_{A}^{1}=~\lambda_{A}^{4}=~\lambda_{A}^{b}=0.2t$, $\lambda_{A}^{2}=~\lambda_{A}^{3}=~\lambda_{A}^{g}=-0.2t$ for Figs.~\ref{fig11}(a),~\ref{fig11}(c) and~\ref{fig11}(e-h),
and $\lambda_{A}^{1}=~\lambda_{A}^{b}=0.2t$, $\lambda_{A}^{2}=~\lambda_{A}^{3}=~\lambda_{A}^{4}=~\lambda_{A}^{g}=-0.2t$ for Figs.~\ref{fig11}(b) and~\ref{fig11}(d).
In Fig.~\ref{fig11}(a), the direction of the antiferromagnetic orderings
in the two nearest junction regions remain opposite,
four zero-energy degenerate topologically-protected 0D in-gap states
appear and are localized at the four edge interfaces not the middle interface point of the heterojunction [see Fig.~\ref{fig11}(c)].
The charges of four in-gap states all are $e/2$ regardless of at the armchair and zigzag edges.
In Fig.~\ref{fig11}(b), only the magnetic ordering direction of blue region-1
is opposite to the other three green regions,
there are two zero-energy degenerate 0D in-gap states.
As plotted in Fig.~\ref{fig11}(d), the two in-gap states are located
at the end of the interface between region-1 and region-2, or between region-1 and region-3 where the antiferromagnetic orderings between them are opposite.
There is no in-gap state at the center of the nanoflake
although there the interface orientation changes by 90 degrees.
The square-shaped finite-size $16 \times 25$ central hollows are introduced in Figs.~\ref{fig11}(e) and ~\ref{fig11}(g).
There are eight topologically-protected in-gap states present at the ends of four interfaces of the heterojunction [see Figs.~\ref{fig11}(e) and~\ref{fig11}(g)].
While the size of the central hollows is reduced to $8 \times 13$ in Figs.~\ref{fig11}(f) and ~\ref{fig11}(h),
Eight in-gap interface end states evolve into
four zero-energy and four non-zero-energy in-gap states,
as displayed by the red and blue dots in Fig.~\ref{fig11}(f).
The zero-energy and non-zero-energy in-gap states are localized at
the outer and inner ends of four interfaces of the heterojunction,
respectively [see Fig.~\ref{fig11}(h)].
The deviation from the zero energy for the four inner end states originates from their couplings each other in case of the small central hollow.
With the decreasing of the hollow's size,
the four inner in-gap end states gradually deviate from the zero energy.
Until the central hollows disappear, the four non-zero-energy in-gap states are
completely transformed into the gapped edge states [see Fig.~\ref{fig11}(a)].
That is, the four inner end states couple to each other and then vanish.
This is similar to that two Majorana zero modes couple and then they evolve into two non-zero-energy normal states.\cite{Maj1,Maj2}
It can be concluded that multiple topologically-protected 0D
in-gap interface end states can be induced by multiple region coupling and hollows
as long as the antiferromagnetic ordering directions on both sides of the interfaces are opposite.

 \begin{figure}
	\centering
	\includegraphics[width=8.6cm,angle=0]{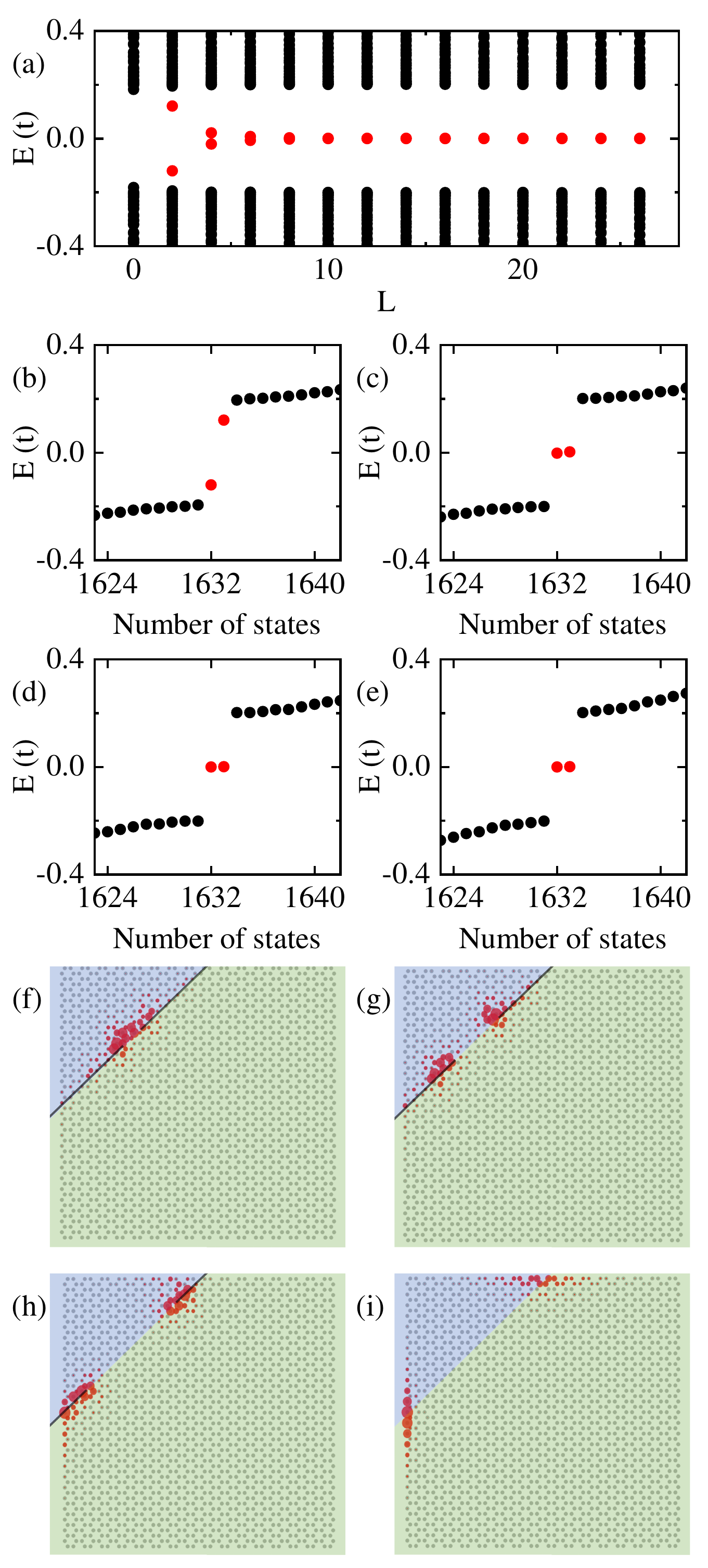}
	\caption{(a) Energy levels of finite-size $32 \times 51$ heterojunction nanoflake consisting of right triangle blue region and pentagon green region
with different antiferromagnetic orderings as a function of coupling length $L$ along inclined interface line.
The coupling strength $\gamma$ equal to $1$ within the coupling length $L$, otherwise $\gamma=0$.
(b-i) Energy levels and wave function distributions
of finite-size heterojunction nanoflake with different coupling lengths $L=2$ for (b) (f), $L=8$ for (c) (g), $L=16$ for (d) (h), and $L=26$ for (e) (i). The area of the red circles in (f-i) are proportional to the charge.
Other parameters (including the antiferromagnetic ordering magnitudes in
the blue and green regions) are the same as those in Fig.~\ref{fig9}(d).
The black lines in (f), (g) and (h) represent the uncoupled interface region.}
	\label{fig12}
\end{figure}

Finally, we construct an irregular interface in a heterojunction by changing region-1 to the shape of a right triangle on the premise of Fig.~\ref{fig11}(d).
In this way, the interface between region-1 and the other three regions
is along neither zigzag nor armchair edge and is more generic.
In Fig.~\ref{fig12}(a), we plot the energy levels $E$ of the finite-size heterojunction nanoflakes as a function of coupling length $L$.
The parameters are the same as those in Fig.~\ref{fig11}(b,d)
except for the uncoupled region $\gamma=0$.
In addition, we plot energy levels and wave function distribution
of in-gap states with different coupling lengths $L=2$ for (b) (f),
$L=8$ for (c) (g), $L=16$ for (d) (h) and $L=26$ for (e) (i).
When $L=0$, the two regions completely separate and there is no state
in the energy gap.
At $L=2$, one can see from Figs.~\ref{fig12}(a) and \ref{fig12}(b)
that the heterojunction system has two in-gap states,
although they deviate heavily from the zero energy.
These two in-gap states are highly overlapping in space as shown in Fig.~\ref{fig12}(f).
As the increasing of $L$, the energy value of the in-gap states
gradually approaches the zero energy.
They almost are zero-energy degenerate when $L=10$.
At $L=8$, there is still a small energy gap between the two in-gap states,
as shown in Fig.~\ref{fig12}(c).
The wave function distributions of two in-gap states have decoupled
in space to form topologically-protected 0D in-gap interface end states
[see Fig.~\ref{fig12}(g)].
At $L=16$, two topologically-protected 0D in-gap interface end states
with zero-energy degenerate are induced in the heterojunction, as shown in Fig.~\ref{fig12}(d). They in space are well localized at the end of the lattices in the heterojunction coupling region in Fig.~\ref{fig12}(h).
At $L=26$, the two parts of regions are completely coupled,
and two topologically-protected 0D in-gap interface end states
with zero energy appear in Fig.~\ref{fig12}(e).
In Fig.~\ref{fig12}(i), the two in-gap states are located in the interface ends between blue region and green region,
which is no different from Fig.~\ref{fig11}(d).
It indicates that even if the interface is not strictly zigzag or armchair boundary, topologically-protected 0D in-gap interface end states can still be induced by the construction of heterojunction.
The appearance of the topologically-protected 0D in-gap interface end states is independent of the shape and size of the two nanoflakes in the heterojunction.
Besides, $L=2$ is enough to induce in-gap interface states,
and $L=10$ is enough to induce zero-energy degenerate in-gap interface end states. The location of the in-gap interface end states
can be adjusted randomly with the coupling regions and coupling length.

%%========================================================================================
\section{\uppercase\expandafter{\romannumeral 4}. summary and discussions}
%%========================================================================================

In summary, we investigate the antiferromagnetic graphene
heterojunction by using the Kane-Mele model.
We find that the helical edge states open the gap by
the antiferromagnetic ordering and there is no the in-gap corner state
for individual antiferromagnetic graphene.
When a heterojunction consists of two antiferromagnetic graphenes
with opposite in-plane antiferromagnetic orderings,
topologically-protected 0D in-gap states can be induced
by the formation of the heterojunction and they locate at the
end of the interface.
We also show that the 0D in-gap interface end states
are robust against magnetic disorder, Anderson disorder and interfacial
magnetic defects.
The origin of the 0D in-gap interface end states is that
for the individual antiferromagnetic graphene,
the effective mass terms of the helical edge states
have the same sign for the sample's four edges,
but they have the opposite signs on the two sides of
the heterojunction with the opposite antiferromagnetic ordering.
Interestingly, the position of the 0D in-gap states
in the heterojunction sample is highly adjustable
by changing the coupling length.
Furthermore, several other heterojunction configurations
are considered including multi-region heterojunction
and irregular interfacial heterojunction.
The multiple 0D in-gap states are induced at the sample edge interface.
Meanwhile, the topologically-protected 0D in-gap interface end states
can freely tunable in the whole sample by changing the coupling region
and length without relying on zigzag or armchair boundaries.

At last, let us discuss whether there exists
a topological index of bulk that is corresponding to the
presence of the 0D in-gap interface end states.
It is well known that there is the bulk-boundary correspondence
in the topological insulator, and
the presence of the gapless edge states is determined by the properties of
the bulk bands.
However, in our antiferromagnetic heterojunction graphene system,
the 0D in-gap states locate at the end of
the interface of two opposite antiferromagnetic graphenes.
So if there exists the bulk-boundary correspondence for
the 0D interface end states,
the topological index will relate to the bulk bands of two systems
(two opposite antiferromagnetic graphenes).
This is essentially different from the conventional bulk-boundary correspondence
in the topological insulator, in which
its topological index only relates to the bulk band of one system (the topological insulator).

\begin{acknowledgments}
\section{acknowledgments}
This work was financially supported by the National Natural Science Foundation of China (Grant No. 12074097, and No. 11921005),
Natural Science Foundation of Hebei Province (Grant No. A2020205013),
the Innovation Program for Quantum Science and Technology (2021ZD0302403),
and the Strategic Priority Research Program of Chinese Academy of Sciences (Grant No. XDB28000000).

C.M.M. and Y.H.W. contributed equally to this work.
\end{acknowledgments}

%\bibliography{Heterojunction}

\end{document}